\documentclass[aps,twocolumn,prl,showpacs,preprintnumbers,amsmath,amssymb]{revtex4}

\usepackage{graphicx}
\usepackage{dcolumn}
\usepackage{bm}

\begin{document}

\title{
Dynamic Nuclear Polarization in Silicon Microparticles}\

\author{A. E. Dementyev$^{1}$\footnote{Author to whom correspondence should be addressed. Electronic address:
anatolyd@physics.harvard.edu}}
\author{D. G. Cory$^{2}$}
\author{C. Ramanathan$^{2}$}

\address{${^1}$Francis Bitter Magnet Laboratory, Massachusetts Institute of Technology, Cambridge, MA 02139, USA
\\ $^{2}$Department of Nuclear Science and Engineering, Massachusetts Institute of Technology, Cambridge, MA 02139, USA}

\date{\today}

\begin{abstract}

We report record high $^{29}$Si spin polarization obtained using dynamic nuclear polarization in
microcrystalline silicon powder. Unpaired electrons in this silicon powder are due to dangling bonds in the
amorphous region of this intrinsically heterogeneous sample. $^{29}$Si nuclei in the amorphous region become
polarized by forced electron-nuclear spin flips driven by off-resonant microwave radiation while nuclei in the
crystalline region are polarized by spin diffusion across crystalline boundaries. Hyperpolarized silicon
microparticles have long $T_{1}$ relaxation times and could be used as tracers for magnetic resonance imaging.


\end{abstract}


\maketitle

 Crystalline and amorphous silicon are both of great technological importance. Single-crystal silicon is a
work-horse of the electronic industry while amorphous silicon is widely used for photovoltaic devices. Recently,
several promising silicon-based architectures have been proposed for quantum computing \cite{kane,vrijen,ladd}.
The long coherence times of $^{29}$Si and $^{31}$P nuclear spins in lightly doped silicon crystals make them
attractive candidates for quantum bits (qubits)\cite{abragam,feher}. One of the major limitations of using
nuclear spins as qubits is the highly mixed state of the nuclear spin system at thermal equilibrium
\cite{cory,chuang,warren}. Several methods can be used to hyperpolarize nuclear spin systems. For instance,
optical pumping techniques have been very successful in studies of electron spin physics in gallium arsenide
quantum wells and quantum dots \cite{barrett,bracker} but so far have produced comparatively low nuclear spin
polarization in silicon \cite{lampel,bagraev,verhulst}. Microwave induced dynamic nuclear polarization (DNP) is
a powerful technique, where an enhancement of several orders of magnitude of the nuclear spin polarization can
be generated by microwave irradiation of paramagnetic centers coupled to the nuclear spin system through the
hyperfine interaction \cite{goldman,griffin}. Since the discovery of DNP in 1953 \cite{overhauser,slichter},
there have been several reports of DNP in silicon \cite{abragamSi,wind,wenckebach}, but all of them were done
either at high temperatures or low magnetic fields limiting the $^{29}$Si spin polarization to a few percent. It
is imperative for the emerging fields of spintronics \cite{wolfe} and nanotechnology to develop methods to
achieve high $^{29}$Si spin polarization in both bulk and nanoscale samples.

Here, we report the first DNP experiments in undoped silicon microparticles. The amorphous region of these
particles has a very high concentration of structural defects - dangling bonds \cite{brodsky}. $^{29}$Si nuclei
in this region are polarized by microwave irradiation of the hyperfine coupled electron -nuclear spin system.
The density of paramagnetic impurities in the crystalline cores is low which results in the long nuclear
relaxation times. Nuclei in those regions are polarized by nuclear spin diffusion \cite{bloembergen} across
crystalline boundaries. Several orders of magnitude enhancement of the NMR signal combined with the long T$_1$
relaxation time makes this a valuable technique for hyperpolarizing silicon nanoparticles, with potential
applications in medical imaging \cite{marcus,nanobioimaging} and NMR studies of nanoscale systems (e.g. silicon
nanowires).

\begin{figure}
\scalebox{0.35}{\includegraphics{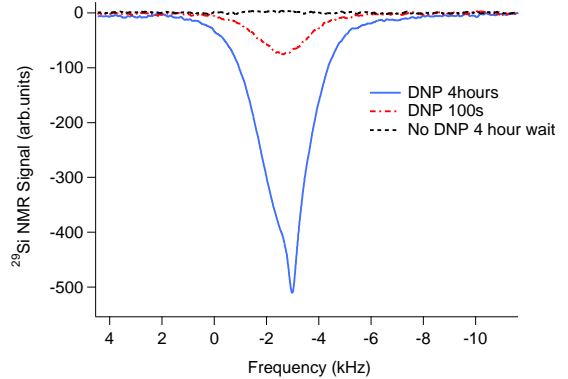}} \caption{\label{mylabel} $^{29}$Si NMR spectra acquired at 1.4K for
different microwave irradiation times in B=2.35T (f$_0$=19.89MHz). The spectrum acquired without the microwave
irradiation is shown using the dashed line.}
\end{figure}

Our sample is a polycrystalline powder obtained from Alfa Aesar. It is 99.999\% pure with particle sizes
distributed between 1 and 5 $\mu$m. According to X-ray diffraction analysis of this sample 80\%  of the sample
is amorphous while 20\% is crystalline and individual crystallites are larger than 200nm. Our DNP experiments
were performed at 1.4 K in a 2.35 T superconducting NMR magnet.  An Oxford NMR Spectrostat was operated in
single-shot mode to cool the sample down to 1.4 K. The microwave source was a 90 mW Gunn diode source
(Millitech). The NMR spectrometer used was a Bruker Avance system with a home-built probe, containing a horn
antenna for the microwave irradiation of the sample, and a solenoidal NMR rf coil, that were in direct contact
with pumped liquid helium. Details of a similar probe design have been described elsewhere\cite{dnpprobe}.

Figure 1 shows the $^{29}$Si NMR spectra acquired without and with microwave irradiation at 66.25 GHz for
different irradiation times. For short irradiation times the NMR spectrum is inhomogeneously broadened with a 2
kHz linewidth (full width at half of a maximum height) due to the distribution of distances between nuclear and
electron spins in the amorphous region. The sharp feature in the spectrum is due to nuclei in the crystalline
cores and appears for microwave irradiation times longer than an hour. The 300 Hz linewidth of this feature is
consistent with previous NMR measurements on silicon crystalline powder samples \cite{dementyev} and can be
explained by a combination of dipole-dipole interaction between $^{29}$Si nuclei arranged in the crystalline
lattice and magnetic susceptibility broadening. The average DNP enhancement of the nuclear polarization for the
amorphous phase inferred from Fig.1 is 150, corresponding to a $^{29}$Si polarization of 5\%. We believe this to
be the highest $^{29}$Si polarization yet achieved. Nuclear polarization of the crystalline region after 4 h of
DNP is found to be $\sim$ 2\%. This estimate comes from the ratio of the integrated NMR signals from amorphous
and crystalline regions and takes into account that crystalline region is only 20\% of the sample.

\begin{figure}
\scalebox{0.40}{\includegraphics{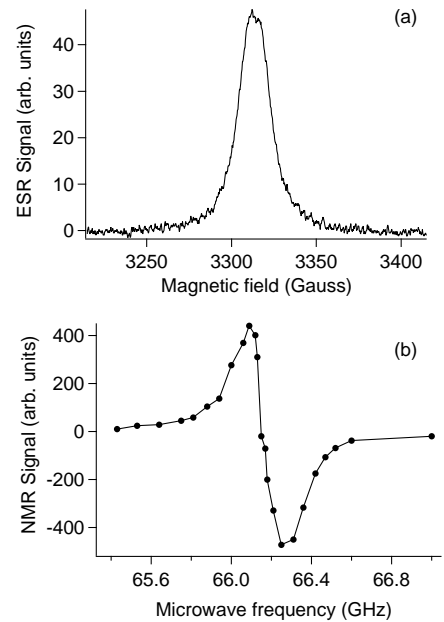}} \caption{\label{mylabel}(a) The ESR spectrum acquired at 3.6 K and
9.27 GHz. (b) The dependence of the NMR signal amplitude on microwave frequency acquired at 2.35T and 1.4K for a
microwave irradiation time of 200 s.}
\end{figure}

We performed ESR studies to characterize the nature of the paramagnetic centers in our sample. The g-factor
value $g = 2.006$, obtained from our ESR spectra is consistent with earlier measurements in amorphous silicon
\cite{brodsky} that identified the defects as dangling bonds. Figure 2(a) shows the low temperature ESR spectrum
which was acquired in the rapid-passage regime with the spectrometer tuned to dispersive response. The electron
spin T$_1$ relaxation time was estimated to be approximately 30 $\mu$s using the rapid-passage phase method
\cite{cullis}.

Figure 2(b) shows the dependence of the NMR signal amplitude on the microwave frequency for a microwave
irradiation time of 200 s acquired at 2.35T and 1.4K. It is characteristic of the DNP thermal mixing mechanism.
This is a typical DNP mechanism in insulators with a high concentration of paramagnetic impurities at low
temperatures \cite{goldman,griffin,farrar}. Microwave irradiation just below the ESR central frequency cools the
electron spin dipolar reservoir while microwave irradiation just above the ESR central frequency creates a
negative temperature for the electron spin dipolar reservoir. Since the ESR linewidth is larger than the nuclear
Larmor frequency, the electron spin flip-flop interaction mediated by nuclear spin flips equalizes the electron
spin dipolar temperature with the nuclear spin temperature.

\begin{figure}
\scalebox{0.4}{\includegraphics{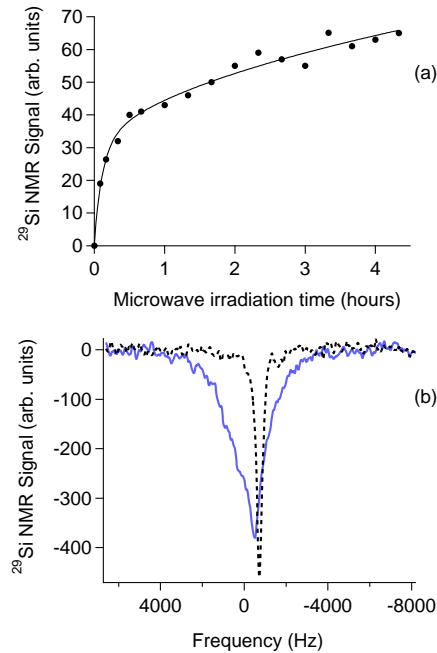}} \caption{\label{mylabel}(a) The dependence of the NMR signal
amplitude on microwave irradiation time. The solid line is to guide the eye. (b) Two $^{29}$Si NMR spectra, the
solid line is the spectrum acquired after 4 h of DNP at 1.4K with a 5$^\circ$ tipping pulse and the dashed line
is the spectrum acquired after the microwave irradiation was turned off and the sample was warmed up to 240K (a
90$^\circ$ tipping pulse was used).}
\end{figure}

The microwave irradiation time dependence of the NMR signal amplitude is shown in Fig. 3(a). There are two
distinct time scales. Within the first hour $^{29}$Si nuclei in the amorphous region are hyperpolarized through
the thermal mixing mechanism. Beyond the first hour the NMR signal continues to increase because of spin
diffusion into the crystalline cores. There is still an increase of the signal for times after 4 h. This is
consistent with very slow spin diffusion in the network of naturally abundant $^{29}$Si spins mediated by XY
terms of the dipolar interaction:

\begin{equation}
\frac{{\mathcal H}_{d}}{\hbar}=\sum_ {j>i}a_{ij} \Big\{I_{z_{i}}I_{z_{j}}-\frac{1}{2}\left(I_{x_{i}}I_{x_{j}} +
I_{y_{i}}I_{y_{j}}\right)\Big\}, \label{eqDipH}
\end{equation}
where $a_{ij}=\frac{(^{29}\gamma)^{2}\hbar}{r_{ij}^{3}} [1-3\cos^{2}\theta_{ij}]$ ($^{29}\gamma$ is the
gyromagnetic ratio for $^{29}$Si). The vector between spins i and j, ${\vec r_{ij}}$, satisfies ${\vec
r_{ij}}\cdot{\hat z} =  r_{ij}\cos\theta_{ij}$. Magnetization transport is described by the diffusion equation:
\begin{equation}
\frac{\partial M}{\partial t}=D\,\Delta M \label{eqDiff}
\end{equation}
where $D\cong\frac{b^2}{50T_{2}}$ is the diffusion constant \cite{bloembergen}, $T_{2}$ is the nuclear
transverse relaxation time and $b$ is the average distance between neighboring $^{29}$Si nuclei. Assuming in our
case $T_{2}\approx 5.6ms$ \cite{dementyev} and $b\approx 0.55n^{-1/3}\approx 0.41$nm \cite{hoch}, where $n$ is
the density of $^{29}$Si nuclei,  we estimate $D\cong 6\times10^{-15}$ cm$^2$/s.

Once the nuclear polarization builds up inside the crystalline cores of the particles it can be stored for a
time on the order of the nuclear T$_1$ relaxation time. For example, Fig 3(b) shows two $^{29}$Si NMR spectra,
one (solid line) was acquired after 4 h of DNP at 1.4K with a short (5$^\circ$) NMR pulse, while the other one
(dashed line) was acquired after the microwave irradiation was turned off and the sample was slowly warmed up to
240K. The broad line from the amorphous phase quickly disappeared due to the fast relaxation rate of the
amorphous domain and only the narrow component from the crystalline cores survived after an hour long warming up
procedure. The resulted $^{29}$Si polarization is about 3000 times higher than the equilibrium polarization at
240K.

\begin{figure}
\scalebox{0.34}{\includegraphics{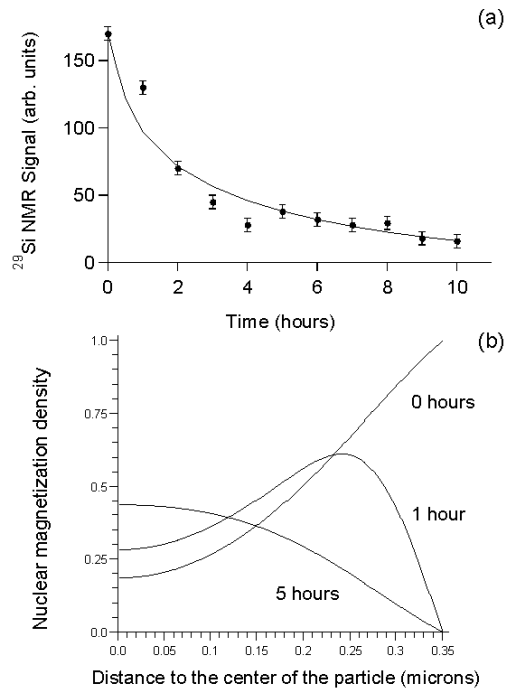}} \caption{\label{mylabel}(a) The decay of the NMR signal from
crystalline silicon after DNP for 5 h at 1.4K. We used 15$^\circ$ NMR pulses to monitor the signal decay. The
solid line is the calculation of the decay due to the spin diffusion of nuclear polarization to the surface. (b)
Simulated nuclear spin magnetization density profiles in a spherical particle with a radius of 350nm
corresponding to different time points in (a): right after DNP (labeled 0 hours), 1 hour and 5 hours after DNP.}
\end{figure}

In order to estimate the $^{29}$Si nuclear T$_1$ in the crystalline part of the sample, we prepared nuclear spin
polarization by DNP at 1.4K for 5 h, allowing for nuclear spin diffusion to penetrate inside the crystalline
cores. Afterwards, the microwave radiation was turned off and the cryostat was warmed up to 4.2K to enable the
continuous flow operation. Figure 4(a) shows the decay of the NMR signal amplitude from the crystalline cores
together with a simulation (solid line) using the diffusion equation (Eq.\,\ref{eqDiff}) for a spherical
particle with a radius of 350nm. In combination with the large initial polarization, it is the long T$_1$
observed here that underlies the potential use of these particles as tracers.

In the simulation we set the magnetization density on the surface of the particle to a finite value for the 5 h
DNP period and to a zero right afterwards. The nuclear spin magnetization density profiles obtained using this
simulation are shown in Fig. 4(b). The agreement between the experiment and the simulation is satisfactory given
the simplicity of the model. The decay of the spin magnetization in this model depends on the particle size and
the nuclear spin diffusion coefficient. Although the decay is nonexponential, we calculate the nuclear
spin-lattice relaxation time T$_1$ as the time when magnetization decays to 1/$e$ of the initial value. It
scales as T$_1\propto\frac{R^2}{D}$, where R is the radius of a particle. There is an interplay between the
initial magnetization and the T$_1$ on one side, and the isotopic abundance of $^{29}$Si on the other. High
$^{29}$Si concentration leads to a larger initial magnetization but also to a larger nuclear spin diffusion
coefficient and thus a shorter T$_1$. The result is that isotopic enrichment reduces the storage time. Figure 5
shows simulations of the $^{29}$Si magnetization decay due to spin diffusion and instantaneous relaxation at the
surface for different particle sizes and for naturally abundant (4.67\%) and enriched (100\%) $^{29}$Si. The
ratio of spin diffusion coefficients for enriched and naturally abundant $^{29}$Si is about 10 \cite{silbey}.
Although the initial magnetization density is 21.4 times larger for enriched silicon, the magnetization decay is
also much faster due to faster diffusion and there is always a crossing between the two curves at long times.

\begin{figure}
\scalebox{0.39}{\includegraphics{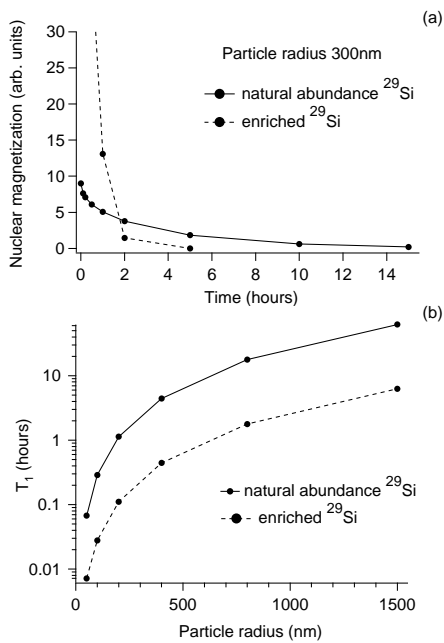}} \caption{\label{mylabel}(a) Simulations of the $^{29}$Si
magnetization decay due to spin diffusion and instantaneous relaxation at the surface for particles with 300nm
radius. (b) Simulated T$_1$ relaxation time as a function of the particle size. Solid line is for the naturally
abundant $^{29}$Si and dashed line is for the $^{29}$Si enriched sample. }
\end{figure}

In summary, we have extended previous DNP experiments in silicon to lower temperature and higher magnetic field
and have achieved record high $^{29}$Si spin polarization in silicon microparticles. This technique might also
be used for polarizing nuclei in thin films of crystalline silicon if a layer of amorphous silicon is created on
top (e.g., by ion implantation). Nuclear spin diffusion in $^{29}$Si abundant crystals is significantly faster
which will make it possible to polarize thicker films. The decay of nuclear spin polarization in the crystalline
part of the sample is due to nuclear spin diffusion to the surface and subsequent relaxation by paramagnetic
impurities. The use of hyperpolarized nanoparticles as tracers depends on both the size and time scales of
interest in a given application.  While moderately sized particles can retain their polarization for a
relatively long time, the relaxation becomes very fast for the smallest nanoparticles.

A.E.D thanks C.M. Marcus, R.L. Walsworth, S.E. Barrett, and J. Baugh for helpful discussions, and S.A. Speakman
for the X-ray diffraction analysis of our sample. This work was supported in part by the National Security
Agency (NSA) under Army Research Office (ARO) contract number DAAD190310125, the NSF and DARPA DSO.

\end{document}